\documentclass[aps,pra,english,showpacs,superscriptaddress,twocolumn]{revtex4}
\usepackage{latexsym}
\usepackage{bm}
\usepackage{babel}
\usepackage[T1]{fontenc}
\usepackage[latin1]{inputenc}
\usepackage{mathrsfs}
\usepackage{amsmath, amssymb}
\usepackage{graphicx}
\usepackage{subfigure}

\begin{document}
\title{Spin squeezing and precision probing with light and samples of atoms in
the gaussian approximation}
\author{Lars Bojer Madsen}
\affiliation{Department of Physics and Astronomy, University of Aarhus,
8000 {\AA}rhus C, Denmark}
\author{Klaus M{\o}lmer}
\affiliation{Danish National Research Foundation Center for Quantum Optics
and Department of Physics and Astronomy, University of Aarhus, 8000 {\AA}rhus C,
Denmark}

\begin{abstract}
We consider an ensemble of trapped atoms interacting with a
continuous wave laser field. For sufficiently polarized atoms and
for a polarized light field, we may approximate the non-classical
components of the collective spin angular momentum operator for
the atoms and the Stokes vectors of the field by effective
position and momentum variables for which we assume a gaussian
state. Within this approximation, we present a theory for the
squeezing of the atomic spin by polarization rotation measurements
on the probe light. We derive analytical expressions for the
squeezing with and without inclusion of the noise effects
introduced by atomic decay and by photon absorption. The theory is
readily adapted to the case of inhomogeneous light-atom coupling
[A. Kuzmich and T.A.B. Kennedy, Phys. Rev. Lett. {\bf 92}, 030407
(2004)]. As a special case, we show how to formulate the theory
for an optically thick sample by slicing the gas into pieces each
having only small photon absorption probability. Our analysis of a
realistic probing and measurement scheme shows that it is the
maximally squeezed component of the atomic gas that determines the
accuracy of the measurement.

\end{abstract}

\maketitle

\section{Introduction}
\label{sec:Introduction} With spin squeezed atomic ensembles,
i.e., samples where the variance of one of the angular momentum
(spin) components is reduced compared with the coherent state
value, one has the possibility to measure certain atomic and/or
classical parameters beyond the precision set by the standard
quantum noise. Recent examples where this possibility was
exploited include studies on magnetometry with collective atomic
spins \cite{geremia03,Moelmer04}. The central feature in those
works is the entanglement of collective continuous light-atom
variables. This entanglement can be created by the free-space
interaction between a trapped polarized atomic sample and an
appropriately polarized propagating laser beam with photon energy
adjusted to the energy spacing between the atomic energy levels
\cite{duan00,Julsgaard01}. The probing of the atomic ensemble with
the light field squeezes the atomic observable (the atomic spin)
and enables an improved measurement, e.g., of a magnetic field.
The underlying squeezing of the collective atomic spin variable
was dealt with in a series of papers (see, e.g., Refs.
\cite{Kuzmich98,Takahashi99,Kuzmich99,Bouchoule02,Muller04,Thomsen02,Kuzmich04},
and references therein)  including investigations of quantum
non-demolition feedback schemes \cite{Thomsen02,Geremia04}, and a
study of the case of inhomogeneous light-atom coupling
\cite{Kuzmich04}. In the present work, we follow the lines of
Refs.~\cite{Kraus03,hammerer,Moelmer04}, and investigate the
spin-squeezing of continuous variable quantum systems in the
approximation where the atomic and photonic degrees of freedom are
described by a gaussian state. To this end we will use that the
gaussian state is fully characterized by its expectation value
vector and its covariance matrix and we will use that explicit
formulae exist for the time evolution of the system {\it and} for
the quantum state reduction under measurements, see, e.g., Refs.
\cite{GiedkeCirac,Fiurasek02,EisertPlenio} and references therein.
In particular, the fact that the measurements are explicitly
accounted for in the gaussian approximation is a strength of the
present theory.

In the development of the theory, we shall consider a continuous
wave (cw) beam of light passing through a cloud of trapped atoms.
In the Schr\"{o}dinger picture we have an explicit update formula
for the quantum state conditioned on the outcome of measurements
carried out on a quantum system, but a light beam is a multi-mode
field with an infinite dimensional Hilbert space, in which a
complete description of the quantum state is normally
prohibitively complicated. The quantum mechanical description of
cw optical fields is often formulated in terms of temporal
correlation functions or the noise power spectrum of field
operators in the Heisenberg picture, which is, however, not a
convenient formulation, when the field is being monitored
continuously in time. When we restrict ourselves to gaussian
states, however, it is possible to describe the field in the
Schr\"{o}dinger picture and to dynamically evolve the combined
quantum state of the interacting light field and atomic system.

The paper is organized as follows. In Sec.~\ref{sec:coll}, we
derive the Hamiltonian for the collective atom-light coupling. In
Sec.~\ref{sec:dyn}, we describe dynamics and measurements in the
gaussian approximation and provide update formulae for the
covariance matrix and for the expectation value vector. In
Sec.~\ref{sec:hom}, we present fully analytical results for
spin-squeezing of an atomic gas for a homogeneous light-atom
coupling, and small photon absorption probability and atomic decay
rate. In Sec.~\ref{sec:inhom}, we describe how to handle the case
of inhomogeneous light-atom coupling. In Sec.~\ref{sec:thin} we
treat the case of an optically thin gas, i.e., small photon
absorption, and we obtain analytical results. In
Sec.~\ref{sec:thick}, we investigate the case of an optically
thick gas. In Sec.~\ref{sec:probing}, we show that the maximally
squeezed component of the gas will set the limit for the precision
in a given measurement. Sec.~\ref{sec:conclusions} briefly
summarizes the results and concludes the paper.

\section{Collective atom-light coupling}
\label{sec:coll}

To describe the atom-light coupling,  we imagine that the beam is
split up into short segments of duration $\tau$ and corresponding
length $L=c\tau$. These beam segments are chosen so short, that
the field in a single segment can be treated as a single mode, and
that the state of an atom interacting with the field does not
change appreciably during time $\tau$, so that the evolution of
the atomic system is obtained by sequential interaction with light
segments. Since we are interested in modelling a cw coherent beam
with constant intensity, we assume a mode function for each
segment of the field which is constant on a length $L$ and within
the transverse area $A$, i.e., the quantization of the field
energy $LA\varepsilon_0E^2=N_{\text{ph}}\hbar\omega$ yields the
relation between the electric field amplitude and the photon
number in the segment of the field,
$E=\sqrt{N_{\text{ph}}}\sqrt{\frac{\hbar\omega}{LA\varepsilon_0}}$.
In the scheme for spin squeezing, we consider a light beam
linearly polarized along the $x$ direction and propagating in the
$y$ direction. The polarization can be decomposed in two
polarization components with opposite circular polarization with
respect to the quantization axis $z$. These two components
interact differently with the atoms because of the selection rules
of the optical dipole transition. Imagine atoms with a ground ($|
g\rangle$) and an excited ($| e \rangle$) state with $J=1/2$,
interacting with the $\sigma^+$ and $\sigma^-$ components of the
light field on the
$|g_{-1/2}\rangle\leftrightarrow|e_{1/2}\rangle$ and
$|g_{1/2}\rangle\leftrightarrow|e_{-1/2}\rangle$ transitions,
respectively. The interaction Hamiltonian between a collection of
$N_\text{at}$ atoms, enumerated with the index $i$ and the two
quantized fields thus writes
\begin{eqnarray}\label{hamil1}
H &= & \sum_{i=1}^{N_\text{at}}
\left( \hbar g a_+|e_{1/2,i} \rangle \langle g_{-1/2,i}| + h.c. \right. \nonumber \\
  & + & \left. \hbar  g a_-|e_{-1/2,i}\rangle\langle g_{1/2,i}| + h.c.
\right),
\end{eqnarray}
with $\hbar g=-d E_0$, $d$ the atomic dipole moment on the
relevant transition, and $E_0 = \sqrt{\hbar
\omega/LA\varepsilon_0}$ the 'field per photon', identified above.
We assume that the fields are frequency detuned by an amount
$\Delta$ with respect to the atomic resonance. In the limit where
$g \sqrt{N_{\text{ph}}} \ll \Delta$ the atoms are not excited by
the fields, and the dynamics is entirely associated with the
light-induced energy shifts of the ground states. Adiabatic
elimination of the upper states then leads to the effective
Hamiltonian
\begin{eqnarray}\label{hamil2} H  &= & \sum_{i=1}^{N_\text{at}}
\frac{\hbar g^2}{\Delta}
\left( a^{\dagger}_+a_+|g_{-1/2,i} \rangle \langle g_{-1/2,i}| + \right. \nonumber \\
 & + &  \left.  a^{\dagger}_-a_- |g_{1/2,i} \rangle \langle g_{1/2,i}| \right)   ,
\end{eqnarray}
which applies for the duration $\tau$ for which the field overlaps
the atomic system. The photon field is suitably described by a
Stokes vector formalism, with a macroscopic value of the component
$\langle S_x\rangle = \hbar N_{\text{ph}}/2$, and where the $S_z$
operator yields the difference between the number of photons with
the two circular polarizations, $S_z=\hbar (a_+^{\dagger}a_+ -
a_-^{\dagger}a_-)/2$, and $S_y$ yields the difference between the
number of photons polarized at $45^{\circ}$and $135^{\circ}$, with
respect to the $z$ axis, respectively. The Stokes vector
components obey the commutator relations of a fictitious spin, and
the associated quantum mechanical uncertainty relation on $S_y$
and $S_z$, $\text{Var}(S_y) \text{Var}(S_z) = | \langle \hbar S_x
\rangle |^2/4$, are in precise correspondence with the binomial
distribution of the linearly polarized photons onto the other sets
of orthogonal polarization directions. We introduce the effective
cartesian coordinates \begin{equation} \label{eff-ph}
(x_{\text{ph}},p_{\text{ph}})=\left( \frac{S_y}{ \sqrt{|\langle
\hbar S_x\rangle|}} , \frac{S_z}{ \sqrt{|\langle \hbar
S_x\rangle|}} \right), \end{equation} with the standard commutator
$[x_{\text{ph}},p_{\text{ph}}]=i$ and resulting uncertainty
relation, which is minimized in the initial state, implying that
this state is a gaussian state, i.e., its Wigner function is a
gaussian function of the phase space coordinates.

The atomic ensemble is initially prepared with all $N_{\text{at}}$
atoms in a superposition
$(|g_{-1/2}\rangle+|g_{1/2}\rangle)/\sqrt{2}$ of the two ground
states with respect to the quantization axis $z$, i.e., the total
state of the atoms is initially given by $\left(
(|g_{-1/2}\rangle+|g_{1/2}\rangle)/\sqrt{2}
\right)^{N_\text{at}}$. In this state,  the system of two-level
atoms is described by a collective spin vector, where the
component along the $x$-direction attains the macroscopic value
$\langle J_x\rangle = \hbar N_{\text{at}}/2$, and where the
collective spin along the $z$-axis, $J_z$, represents the
population difference of the $|g_{\pm 1/2}\rangle$ states. As for
the photons, the quantum mechanical uncertainty relation for the
collective spin components of the atomic state corresponds exactly
to the binomial distribution of the atoms on the two ground
states, and also here it is convenient to introduce cartesian
coordinates
\begin{equation}\label{eff-at}
(x_{\text{at}},p_{\text{at}})=\left(\frac{J_y}{\sqrt{|\langle
\hbar J_x\rangle|}},
  \frac{J_z}{\sqrt{|\langle \hbar J_x\rangle|}} \right),
\end{equation}
for which the initial state is a minimum uncertainty gaussian state.

The Hamiltonian (\ref{hamil2}) can be rewritten in terms of the
effective atomic and field variables. First, we note that
$\sum_{i=1}^{N_\text{at}} | g _{\mp 1/2,i} \rangle \langle g_{\mp
1/2,i} | = N_\text{at} /2 \pm J_z/\hbar$ and  that $a^\dagger_\pm
a_\pm = \Phi \tau /2 \pm S_z/\hbar$, where $\Phi$ is the photon
flux. We then insert these expressions in Eq.~(\ref{hamil2}),
leave out a constant energy shift and obtain the effective
interaction Hamiltonian
\begin{equation}\label{hamil3} H\tau = \hbar
\kappa_\tau p_\text{at} p_{\text{ph}}.
\end{equation} We display
the product of $H$ and $\tau$, to expose the effect of the
interaction with the whole segment, and we introduce the effective
coupling 'constant'
\begin{equation} \label{kappa1} \kappa_\tau =
2 \frac{g^2}{\Delta} \sqrt{\frac{| \langle J_x \rangle | }{\hbar}
\frac{|\langle S_x \rangle |}{\hbar} } \tau.
\end{equation}
The free-space coupling constant of light and atoms is small, and
the coarse grained description will be perfectly valid even for
the macroscopic values of $N_\text{ph}= \Phi \tau$ required by our
treatment. The Hamiltonian in Eq.~(\ref{hamil3}) correlates the
atoms and the light fields. It is bilinear in the canonical
variables, and hence preserves the gaussian character of the joint
state of the system \cite{GiedkeCirac}. We have emphasized the
convenience of using gaussian states, because their
Schr\"{o}dinger picture representation is very efficient and
compact. Now, given that every segment of the optical beam becomes
correlated with the atomic sample, as a function of time, the
joint state of the atom and field has to be specified by a larger
and larger number of mean values and second order moments. If no
further interactions take place between quantum systems and the
light after the interaction with the atoms, there is no need to
keep track of the state of the total system. In practice, either
the transmitted light may simply disappear or it may be registered
in a detection process. In the former case, the relevant
description of the remaining system is obtained by a partial trace
over the field state, which produces a new gaussian state of the
atoms. We are interested in the case, where the polarization
rotation of the field is registered, i.e., the observable
$x_{\text{ph}}$ is measured. The effect of measuring one of the
components in a multi-variable gaussian state is effectively to
produce a new gaussian state of the remaining variables as
discussed in detail in Sec.~\ref{sec:dyn}.

\section{Dynamics and measurements in the gaussian approximation including noise} \label{sec:dyn}
Having established the fact that the quantum state of the atoms is
at all times described as a gaussian state, we shall set up the
precise formalism. For the column vector of the four variables
${\bm y}= (x_\text{at},p_\text{at},x_\text{ph},p_\text{ph})^T$
describing the atoms and a single segment of the light beam, the
Heisenberg equations of motion yield
\begin{equation} \label{lintrans}
 {\bm y}(t+\tau) = \mathbf{S_\tau} {\bm y}(t)
\end{equation} with the transformation matrix
\begin{eqnarray}\label{Smatrix} {\mathbf S_\tau} =
\left(%
\begin{array}{cccc}
  1 & 0 & 0 & \kappa_\tau \\
  0 & 1 & 0 & 0 \\
  0 & \kappa_\tau & 1 & 0 \\
  0 & 0 & 0 & 1 \\
\end{array}%
\right).
\end{eqnarray}
From Eq.~(\ref{lintrans}) and the definition of the  covariance
matrix $\gamma_{ij} = 2 \text{Re} \left\langle (y_i - \langle y_i
\rangle ) (y_j - \langle y_j \rangle ) \right\rangle$
\cite{EisertPlenio,GiedkeCirac}, we directly verify that
$\bm{\gamma}$ transforms as
\begin{equation}\label{gammatrans-withoutnoise}
    \bm{\gamma}(t+\tau)  = \mathbf{S_\tau} \bm{\gamma}(t)
    \mathbf{S_\tau}^T,
\end{equation}
due to the atom-light interaction.

In the probing process there is a small probability that the
excited state levels which were adiabatically eliminated from the
interaction Hamiltonian of Eq.~(\ref{hamil3}) will be populated.
If this happens, the subsequent decay to one of the two $m_z= \pm
1/2 $ ground states occurs with the rate
\begin{equation} \label{eta}
    \eta=\Phi \frac{\sigma}{A} \left( \frac{\Gamma^2/4} {\Gamma^2/4 + \Delta^2} \right),
\end{equation}
where $\Gamma$ is the atomic decay width and $\sigma=\lambda^2/( 2
\pi)$ is the resonant photon absorption cross-section. The
consequence of the decay is a loss of spin polarization since a
detection of the fluorescence photons in principle could tell to
which ground state the atom decayed. If every atom has a
probability $\eta_\tau = \eta \tau$ to decay in time $\tau$ with
equal probability into the two ground states, the collective mean
spin vector is reduced by the corresponding factor $\langle
\bm{J}\rangle \rightarrow \langle \bm{J}\rangle (1-\eta_\tau)$.
When the classical $x$-component is reduced this leads to a
reduction with time of the coupling strength $\kappa_\tau \mapsto
\kappa_\tau \sqrt{1- \eta_\tau} $  which was also discussed in
Refs.~\cite{geremia03,hammerer,Moelmer04}. Simultaneously, every
photon on its way through the atomic gas has a probability for
being absorbed \cite{hammerer} \begin{equation} \label{epsilon}
\epsilon = N_\text{at} \frac{\sigma}{A}\left( \frac{\Gamma^2/4}
{\Gamma^2/4 + \Delta^2} \right). \end{equation} This means that
the vector of expectation values evolves as $\langle
\bm{y}(t+\tau) \rangle = \mathbf{L_\tau S_\tau } \langle \bm{y}(t)
\rangle$ with $\mathbf{L_\tau} =
\text{diag}(\sqrt{1-\eta_\tau},\sqrt{1-\eta_\tau},
\sqrt{1-\epsilon}, \sqrt{1-\epsilon})$.

The fraction $\eta_\tau$ of  atoms that have decayed represents a
loss of collective squeezing because its correlation with the
other atoms is lost, whereas it still provides a contribution
$\hbar^2/4$ per atom to the collective spin variance. We may use
the symmetry of the collective spin operator under the exchange of
particles to express the mean value of, e.g., $J_z^2$ as $\langle
J_z^2 \rangle = \frac{\hbar^2}{4} N_\text{at}  + \frac{\hbar^2}{4}
N_\text{at}(N_\text{at} -1) \langle \sigma_z^{(1)} \sigma_z^{(2)}
\rangle$ where we have used that $\langle (\sigma_z^{(i)})^2
\rangle = \langle (\sigma_z^{(1)})^2 \rangle = 1$, and $\langle
\sigma_z^{(i)} \sigma_z^{(j)} \rangle = \langle \sigma_z^{(1)}
\sigma_x^{(2)} \rangle$ for all $i$ and $j$ ($i\neq j$). We may
solve the equation for the correlations between the different
spins
\begin{equation}\label{correl}
    \langle \sigma_z^{(1)} \sigma_z^{(2)} \rangle  = \frac{\langle J_z^2
    \rangle - \frac{\hbar^2}{4} N_\text{at}}{\frac{\hbar^2}{4} N_\text{at}
    (N_\text{at} -1)}.
\end{equation}
During a time interval of duration $\tau$, $\eta_\tau N_\text{at}$
atoms decay by spontaneous emission. This means that $\langle
J_z^2 \rangle \mapsto \langle J_z^{\prime 2} \rangle =
\frac{\hbar^2}{4} N_\text{at} (1-\eta_\tau)  + \frac{\hbar^2}{4}
N_\text{at} (1-\eta_\tau) (N_\text{at} (1-\eta_\tau)-1) \langle
\sigma_z^{(1)} \sigma_z^{(2)} \rangle + \frac{\hbar^2}{4}
\eta_\tau N_\text{at}$, where the last term comes from the atoms
that have decayed. The correlations given by Eq.~(\ref{correl})
are inserted, and for large $N_\text{at}$ we find
\begin{eqnarray}\label{spinreduc}
\langle J_z^2\rangle \rightarrow
\langle J_z^{\prime 2} \rangle &=& (1-\eta_\tau)^2 \langle
J_z^2\rangle + \frac{\hbar^2
N_\text{at}}{4}(1-(1-\eta_\tau)^2) \nonumber \\
& \simeq & (1-\eta_\tau)^2 \langle J_z^2\rangle + \frac{\hbar^2 N_\text{at}}{4} 2 \eta_\tau,
\end{eqnarray}
where the last line follows in the limit of small atomic decay,
$\eta_\tau \ll 1$. To determine the development of the canonical
variables, we also need the behavior of moments of the type
$\langle J_x \rangle$: $\langle J_x \rangle \mapsto \langle
J_x^{\prime} \rangle = (1-\eta_\tau) \langle J_x \rangle$.
Combining this result with Eq.~(\ref{spinreduc}), we find
\begin{equation}\label{psq}
\langle p^2_\text{at} \rangle
\rightarrow \langle p'^2_\text{at} \rangle = (1-\eta_\tau )
\langle p_\text{at}^2 \rangle + \frac{\hbar N_\text{at} 2
\eta_\tau/4}{\langle J'_x \rangle}
\end{equation}
and a similar expression for $x_\text{at}$.

The photons that are absorbed do not contribute to the collective
Stokes vector, and we find by an analysis similar to the above,
that
\begin{eqnarray}\label{stokes} \langle S_z^2\rangle \rightarrow
\langle S'^2_z \rangle &=& (1-\epsilon)^2 \langle S_z^2\rangle +
(\hbar^2 N_\text{ph}/4)\epsilon(1-\epsilon) \nonumber \\ & \simeq&
(1-\epsilon)^2 \langle S_z^2\rangle + (\hbar^2
N_\text{ph}/4)\epsilon \end{eqnarray} in the limit of small
$\epsilon$. For the effective $p_\text{ph}$ variable, we find
\begin{equation} \label{pphsq} \langle p^2_\text{ph} \rangle
\rightarrow \langle p'^2_\text{ph} \rangle = (1-\epsilon ) \langle
p_\text{ph}^2 \rangle + \frac{\hbar N_\text{ph}  \epsilon
/4}{\langle S'_x \rangle}, \end{equation} and a similar expression
for $x_\text{ph}$.

Using Eq.~(\ref{psq}) and Eq.~(\ref{pphsq}) and similar
expressions for the other elements of the covariance matrix,
Eq.~(\ref{gammatrans-withoutnoise}) generalizes to
\begin{equation}\label{gammatrans-withnoise} \bm{\gamma}(t+\tau)=
\mathbf{L_\tau} \mathbf{S_\tau} \bm{\gamma}(t) \mathbf{S}^T_\tau
\mathbf{L_\tau} + \frac{\hbar N_\text{at}}{\langle J_x(t)\rangle}
\mathbf{M_\tau} + \frac{\hbar N_\text{ph}}{2 \langle S_x(t)
\rangle} \mathbf{N} \end{equation} for $\eta_\tau, \epsilon \ll 1$
with $\mathbf {M_\tau}= \text{diag}(\eta_\tau,\eta_\tau,0,0)$, and
$\mathbf{ N} = \text{diag}(0,0,\epsilon,\epsilon)$. The factor
$\hbar N_\text{at}/\langle J_x(t)\rangle$ initially attains the
value 2, and increases by the factor $(1-\eta_\tau)^{-1}$ in each
time step $\tau$. The factor $\hbar N_\text{ph}/( 2 \langle
S_x(t)\rangle)$ is initially unity, and is approximately constant
in time since the light field is continuously renewed by new
segments of the light beam interacting with the atoms. An
exception is the optically thick gas discussed below in
Sec.~\ref{sec:thick}.

We note that the present accumulation of noise is based on the
canonical $x$ and $p$ variables entering the covariance matrix,
and not on the physical spin and Stokes variables for the atoms
and the photons, respectively. As discussed in more detail
elsewhere \cite{Sherson04}, this introduces difficulties in the
limit of large atomic decay probabilities. As long as the
probability for atomic decay is small during the process under
concern, however, the present approach is highly accurate. This is
the regime considered in this work.

In the gaussian approximation, the system is fully characterized
by the vector of expectation values $\langle {\bm y} \rangle$ and
the covariance matrix $\bm{\gamma}$. We probe the system by
measuring the Faraday rotation of the probe field, i.e., by
measuring the field observable $x_\text{ph}$. Since the photon
field is an integral part of the quantum system, this measurement
will change the state of the whole system, and in particular the
covariance matrix of the atoms. We denote the covariance matrix by
\begin{eqnarray}\label{gammadecompose}
\bm{\gamma}= \left(%
\begin{array}{cc}
  \mathbf{A_\gamma} & \mathbf{C_\gamma} \\
  \mathbf{C}^T_\gamma & \mathbf{B_\gamma} \\
\end{array}%
\right),
\end{eqnarray}
where the $2 \times 2$ sub-matrix $\mathbf{A_\gamma}$ is the
covariance matrix for the variables ${\bm y }_1 = (x_\text{at},
p_\text{at})^T$, $\mathbf{B_\gamma}$ is the $2 \times 2$
covariance matrix for ${\bm y}_2 = (x_\text{ph}, p_\text{ph})^T$,
and $\mathbf{C_\gamma}$ is the $2 \times 2$ correlation matrix for
$\bm{y}_1$ and $\bm{y}_2^T$. An instantaneous measurement of
$x_\text{ph}$ then transforms $\mathbf{A_\gamma}$ as
\cite{Fiurasek02,GiedkeCirac,EisertPlenio}
\begin{equation}\label{Agammaupdate}
    \mathbf{A_\gamma} \mapsto  \mathbf{A}'_\gamma = \mathbf{A}_\gamma  -
    \mathbf{C_\gamma}
    (\mathbf{\pi B_\gamma} \mathbf{\pi} )^{-} \mathbf{C}^T_\gamma,
\end{equation}
where $\mathbf{\pi} = \text{diag}(1 ,0)$, and where $()^{-}$
denotes the Moore-Penrose pseudoinverse.

After the measurement, the field part has disappeared, and a new
beam segment is incident on the atoms. This part of the beam is
not yet correlated with the atoms, and it is in the oscillator
ground state, hence the covariance matrix $\bm{\gamma}$ is updated
with $\mathbf{A}'_\gamma$, $\mathbf{C}'_\gamma$ a $2 \times 2$
matrix of zeros, and $\mathbf{B}'_\gamma= \text{diag}(1, 1)$
before the next application of the transformation of Eq.~(17).

Unlike the covariance matrix update, which is independent of the
value actually measured in the optical detection, the vector
$\langle \bm{y} \rangle$ of expectation values will change in a
stochastic manner depending on the outcome of these measurements.
The outcome of the measurement on $x_\text{ph}$ after the
interaction with the atoms is random, and  the actual measurement
changes the expectation value of all other observables due to the
correlations represented by the covariance matrix. Let $\chi$
denote the difference between the measurement outcome and the
expectation value of $x_\text{ph}$, i.e., a gaussian random
variable with mean value zero and variance 1/2. The change of
$\langle \bm{y}_1\rangle$ due to the measurement is now given by:
\begin{equation}\label{expectationvalueupdate}
 \langle {\bm y}_1
\rangle \mapsto  \langle {\bm y}'_1 \rangle  = \langle {\bm y}_1
\rangle + \mathbf{C_\gamma} (\mathbf{\pi B
\pi})^{-}(\chi,\cdot)^T,
\end{equation}
where we use that $(\mathbf \pi B \pi)^{-} =
\text{diag}(B(1,1)^{-1}, 0)$, and hence the second entrance in the
vector $(\chi,\cdot)$ need not be specified.

The gaussian state of the system is propagated in time by repeated
use of Eq.~(17) and the measurement update formulae
(\ref{Agammaupdate})-(\ref{expectationvalueupdate}). This
evolution is readily implemented numerically, and the expectation
value and our uncertainty about, e.g.,  the value of the squeezed
$p_\text{at}$ variable of the atoms are given by the second
entrance in the vector of expectation values $\langle y_2 \rangle
= \langle p_\text{at} \rangle$ and the covariance matrix element
$A_\gamma(2,2) = 2 \text{Var}(p_\text{at})$.

We conclude this section by noting that if one associates with the
precise measurement of $x_{ph}$ an infinite variance of $p_{at}$
and a total loss of correlations between $p_{at}$ and the other
variables due to Heisenberg's uncertainty relation, the
Moore-Penrose pseudoinverse can be written as a normal inverse of
the covariance matrix, $(\pi \mathbf{B}\pi)^- =
$diag$(B(1,1),\infty)^{-1}$. Equations (\ref{Agammaupdate}) and
(\ref{expectationvalueupdate}) are then equivalent with the
results for the estimation of classical gaussian random variables
derived, e.g., in Ref.~\cite{Maybeck}.

\section{Homogenous light-atom coupling}
\label{sec:hom}
The time evolution of the atomic $p_\text{at}$
variable is completely determined by the update formulae for the
covariance matrix (\ref{gammatrans-withnoise}) and the measurement
update formula (\ref{Agammaupdate}). In the limit of infinitesimal
time steps, these formulae translate into differential equations,
and we obtain the following equations for the variance of
$p_\text{at} (\propto J_z)$ \begin{equation}\label{varpnospon}
\frac{d}{dt} \text{Var}(p_\text{at}) = - 2 \kappa^2 \left(
\text{Var}(p_\text{at}) \right)^2,
\end{equation}
and
\begin{equation} \label{varpspon}
\frac{d}{dt} \text{Var}(p_\text{at}) = - 2 \kappa^2 (1-\epsilon)
e^{- \eta t} \text{Var}(p_\text{at}) ^2 - \eta \text{Var}(p_\text{at})+ \eta e^{\eta t},
\end{equation}
corresponding to the cases where atomic decay and photon absorption are neglected and
included, respectively. Here the light-atom coupling $\kappa$ is given by
\begin{equation}\label{kappa0} \kappa^2= N_\text{at} \Phi \left( \frac{\chi}{ \Delta }\right)^2,
\end{equation}
with $ \chi = g^2  \tau = \frac{d^2 \hbar \omega}{A c \epsilon_o
\hbar^2}$. Equation (\ref{varpnospon}) is readily solved by
separating the variables, and we obtain
\begin{equation}\label{dp}
    \text{Var}(p_\text{at}) = \frac{1}{2 \kappa^2 t +
    1/ \text{Var}(p_\text{at,0})},
\end{equation}
where $\text{Var}(p_\text{at,0})= 1/2$ is the variance of the initial minimum uncertainty state.

To solve (\ref{varpspon}), we introduce the change of variable
$\widetilde{\text{Var}(p_\text{at})} = e^{-\eta t}
\text{Var}(p_\text{at})$, and obtain
\begin{equation}\label{helpsol} \frac{d}{dt}
\widetilde{\text{Var}(p_\text{at})} = - 2 \kappa^2(1-\epsilon)
\widetilde{\text{Var}( p_\text{at})}^2 - 2 \eta
\widetilde{\text{Var}(p_\text{at})} + \eta,
\end{equation}
which is separable. With $\beta=
\sqrt{\frac{\eta}{\kappa^2(1-\epsilon)}
\left(\frac{\eta}{\kappa^2(1-\epsilon)} +2 \right)}$, the solution
of Eq.~(\ref{varpspon}) reads
\begin{widetext} \begin{equation}
\label{dpe} \text{Var}(p_\text{at}) = \frac{\beta}{2} \left(
\frac{\text{Var}(p_\text{at,0}) + \frac{\eta}{2
\kappa^2(1-\epsilon)} + \frac{\beta}{2} + e^{-2\beta
\kappa^2(1-\epsilon) t} \left( \text{Var}(p_\text{at,0}) +
\frac{\eta}{2 \kappa^2(1-\epsilon)} - \frac{\beta}{2}
\right)}{\text{Var}(p_\text{at,0}) + \frac{\eta}{2
\kappa^2(1-\epsilon)} + \frac{\beta}{2} - e^{-2\beta
\kappa^2(1-\epsilon) t} \left(\text{Var}(p_\text{at,0}) +
\frac{\eta}{2 \kappa^2(1-\epsilon)} - \frac{\beta}{2}\right)}
\right) e^{\eta t} -\frac{\eta}{2 \kappa^2(1-\epsilon)} e^{\eta
t}.
\end{equation}
\end{widetext}
\begin{figure}
\begin{center}
\includegraphics[scale=0.45]{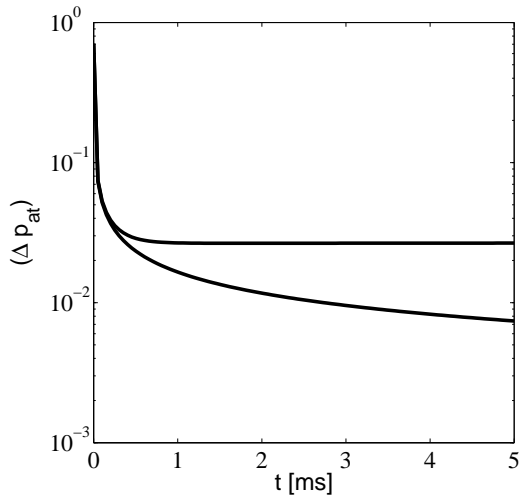}
    \end{center}
    \caption{Uncertainty of $p_\text{at}$
    as function of time.
    The effective coupling is $\kappa^2 =
    1.83 \times 10^{6}$\,s$^{-1}$.
    The lower curve is without inclusion of atomic decay, and
    the upper curve includes atomic decay with a
    rate $\eta =1.7577$\,s$^{-1}$ and photon absorption with
    $\epsilon=0.028$. These values correspond for example to a 2 mm$^2$ interaction
    area, $2 \times 10^{12}$ atoms, $5 \times 10^{14}$ photons s$^{-1}$,
    10\,GHz detuning, and 852\,nm light,
    appropriate for the
    $^{133}$Cs($6S_{1/2}(F=4) - 6P_{1/2}(F=5))$ transition with
    decay rate $3.1 \times 10^7$\,s$^{-1}$ and corresponding
    atomic dipole moment $d=2.61 \times 10^{-29}$\,Cm. Factors of order unity related
    to the coupling
    matrix elements among different states of the actual
    Zeeman substructure are omitted.
    }
    \label{fig:fig1}
\end{figure}

Figure \ref{fig:fig1} shows the spin squeezing as a function of
probing time. When atomic decay is not included, the uncertainty
in $p_\text{at}$ is a monotonically decreasing function with time.
When decays are included, a minimum is reached whereafter the
degree of squeezing starts to decrease. On the time scale of the
figure, which is chosen to reflect realistic experimental time
scales, the increase in $\text{Var}(p_\text{at})$ is hardly
visible. From Eq.~(\ref{varpspon}), we find that the minimum in
the variance occurs at the instant of time
\begin{eqnarray} \label{pmingen}
t_\text{min} &=& \frac{1}{2 \beta \kappa^2(1-\epsilon)}
\\ \nonumber
&\times& \ln \left( \frac{\text{Var}(p_\text{at,0}) +
\frac{\eta}{2 \kappa^2(1-\epsilon)} -
\frac{\beta}{2}}{\text{Var}(p_\text{at,0}) + \frac{\eta}{2
\kappa^2(1-\epsilon)} + \frac{\beta}{2} } \frac{4 \beta
\kappa^2(1-\epsilon)}{\eta} \right).
\end{eqnarray}
In the typical experimental situation, $\eta/2
\kappa^2(1-\epsilon) \ll 1$ which means that $\beta \simeq \sqrt{2
\eta/(1-\epsilon)}/\kappa$. In this case Eq.~(\ref{pmingen})
simplifies to
\begin{equation}\label{tmintyp}
    t_\text{min} = \frac{1}{2 \sqrt{2 \eta (1-\epsilon)} \kappa} \ln
    \left(\frac{4 \sqrt{2 (1-\epsilon)} \kappa}{ \sqrt{\eta}} \right).
\end{equation}
From Eq.~(\ref{tmintyp}), we see that $t_\text{min}$ decreases for
increasing coupling strength $\kappa$, and for increasing decay
rate $\eta$. Interestingly, the instant of time for the minimum in
the variance is independent of the initial uncertainty in the
atomic variable $p_\text{at}$.

We may now go back to Eq.~(\ref{varpspon}) and evaluate the value
of the variance at time $t_\text{min}$. In the regime considered
above, and in the figure, we find
\begin{equation}\label{minp}
    \Delta p(t_\text{min}) = \sqrt{ \frac{1}{\kappa}
    \sqrt{\frac{\eta}{2(1-\epsilon)}}}.
\end{equation}
This clearly shows that the higher coupling and the lower decay,
the better spin squeezing. It is the term linear in $\eta$ in
Eq.~(\ref{varpspon}) that is responsible for the 'saturation
effect' in the variance at early times where the exponential is
still close to unity, $e^{\eta t} \simeq 1$.

To specify for a given number of atoms, how many photons we need
to obtain optimal spin squeezing in time $t_\text{min}$ limited
perhaps by other experimental constraints, we express $\eta = \Phi
\frac{\sigma}{A}\frac{\Gamma^2}{\Delta^2}$, $\kappa^2 = \Phi
\frac{\sigma}{A} \frac{\Gamma^2}{\Delta^2} N_\text{at}$, and
insert in Eq.~(\ref{tmintyp}). The slow logarithmic dependence and
factors of order unity can be neglected, and we can introduce
$\epsilon$ via the relation $\epsilon = N_\text{at}
\frac{\sigma}{A} \frac{\Gamma^2}{ \Delta^2}$, and find $\Phi
t_\text{min} \simeq \frac{1}{\epsilon} \sqrt{\frac{A}{\sigma}
N_\text{at}}$. If we accept photon absorption at the percent
level, we obtain
\begin{equation}\label{limits}
    \Phi t_\text{min} \agt 100 \sqrt{\frac{A}{\sigma}}
    \sqrt{N_\text{at}}.
\end{equation}

In our case, we have $A/\sigma \simeq 1.7 \times 10^{7}$. A
realistic upper limit for $t_\text{min}$ is 1 ms,  and from
Eq.~(\ref{limits}) it then follows that the photon flux should
fulfill
\begin{equation}\label{limit2}  \Phi \agt 10^8
\sqrt{N_\text{at}} \frac{1}{\text{s}}.
\end{equation}

\section{Inhomogeneous light-atom coupling}
\label{sec:inhom}
We now consider two scenarios leading to
inhomogeneous light-atom coupling, a case recently discussed
theoretically in the literature \cite{Kuzmich04}. First, we shall
study the case where the coupling is inhomogeneous as a
consequence of a variation in the intensity of the light beam
across the sample. Second, we shall consider the case of an
optically thick sample where the photon field, and therefore the
coupling, changes through the atomic sample due to absorption.
Both cases are readily handled within the gaussian approximation.

\subsection{Case (a): optically thin sample}
\label{sec:thin}
We consider the case where the atomic gas is
divided into, say $n$, slices each with local light-atom coupling
strength $\kappa_{i}$. The $2n+2$ column vector of gaussian
variables describing the $2n$ collective canonical position and
momentum variables for the atoms, and the 2 collective position
and momentum variables for the photon field then reads
\begin{equation} \label{eq:large-y}  {\bm y}=
(x_{\text{at},1},p_{\text{at},1},\dots,
x_{\text{at},n},p_{\text{at},n}, x_\text{ph},p_\text{ph})^T.
\end{equation}
The generalization of Eq.~(\ref{hamil3}), to the
case with inhomogeneous coupling reads
\begin{equation}\label{Hinhomo}
    H \tau = \hbar \left(\sum_{i=1}^{n} \kappa_{\tau, i} p_{\text{at},i}
    \right) p_\text{ph},
\end{equation}
where the summation index covers the different groups of atoms.

To model the effect of an inhomogeneous coupling of the light  to
the atomic sample, we consider $n=10$ different values of
$\kappa^2$ chosen uniformly in the interval $[
\kappa_0^2(1-\delta); \kappa_0^2(1+\delta) ]$ with $\delta =
\{0,0.1, 0.5\}$. In this way, the effective coupling constant
$\sqrt{\sum_{j=1}^n \kappa_j^2}$ remains constant while the
variance in the coupling constants increases. The values of the
coupling strength could, e.g., differ because of the transverse
intensity profile of the laser beam. As a consequence, the values
of the atomic decay rate $\eta$ (also proportional to intensity)
are different in each slice. The measurement is described by the
method in Sec.~\ref{sec:dyn}, and the propagation is given by a
modification of Eq.~(\ref{gammatrans-withnoise}) \begin{equation}
\label{eq:gamma-inhomo} \mathbf{\gamma}(t+\tau) = \mathbf{L}_\tau
\mathbf{S}_\tau \mathbf{\gamma}_\tau \mathbf{S}_\tau^\dagger
\mathbf{L}_\tau + \mathbf{M}_\tau + \mathbf{N}, \end{equation}
where the $(2n+2) \times (2n+2)$ matrix $\mathbf{S}_\tau$ is
obtained from the time evolution of the system as in
Sec.~\ref{sec:dyn}, and where
\begin{widetext}
$\mathbf{L_\tau} = \text{diag}(\sqrt{1-\eta_{\tau,1}},
\sqrt{1-\eta_{\tau,1}},\dots,\sqrt{1-\eta_{\tau,n}},
\sqrt{1-\eta_{\tau,n}},\sqrt{1-\epsilon}, \sqrt{1-\epsilon})$,
\end{widetext}
$\mathbf{M}_\tau = \hbar\times \text{diag}(\frac{
N_{\text{at},1}\eta_1}{\langle J_{x,1} \rangle }, \frac{
N_{\text{at},1} \eta_1}{\langle J_{x,1} \rangle }, \dots,
\frac{N_{\text{at},n}\eta_n}{\langle J_{x,n} \rangle},\frac{
N_{\text{at},n},\eta_n}{\langle J_{x,n} \rangle }, 0,0)$, and
$\mathbf{N} = \text{diag}(0,0,\dots,0,0,\epsilon, \epsilon)$. For
convenience, we assume that the number of atoms $N_{at,i}$ subject
to a given coupling strength $\kappa_i$ is simply $N_\text{at}/n$.

The atomic covariance matrix now has dimension $(2n \times 2n)$,
and it contains the variances of the atomic observables in each
slice and the correlations between them. Collective observables
are described by linear combinations of the $(x_{at,i},p_{at,i})$
and their variances can be obtained explicitly.

From the Hamiltonian (\ref{Hinhomo}), it is clear, that the probe
field couples to the asymmetric collective variable
$\sum_{i=1}^{n} \kappa_{\tau,i} p_{\text{at},i}$. The
corresponding {\it asymmetric} collective harmonic oscillator
variables involved in the spin squeezing are, accordingly
\begin{equation}
\label{x-p-eff} (X_\text{eff}, P_\text{eff}) = \left(
\frac{\sum_{i=1}^n \kappa_i x_{\text{at},i}}{\sqrt{\sum_{i=1}^n
\kappa_i^2}} ,\frac{\sum_{i=1}^n \kappa_i
p_{\text{at},i}}{\sqrt{\sum_{i=1}^n \kappa_i^2}}  \right),
\end{equation}
The {\it symmetric} collective variables that are usually
considered (see, e.g.,  the discussion in Ref.~\cite{Kuzmich04}
and references therein), are, on the other hand, given by
\begin{equation}\label{symm-col-var}
    (X,P)=\left(\frac{1}{\sqrt{n}}
    \sum_{i=1}^n x_{\text{at},i}, \frac{1}{\sqrt{n}}
    \sum_{i=1}^n p_{\text{at},i} \right),
\end{equation}
and it is interesting to see how these two sets of variables are
connected. A straightforward calculation shows that we may express
the latter variables as
\begin{equation}\label{decompo}
    (X,P)=a ( X_\text{eff}, P_\text{eff}) + b ( X_\bot, P_\bot), \end{equation}
where $(X_\bot,P_\bot)$ are canonical variables which commute with
($X_\text{eff},P_\text{eff}$) and with the interaction Hamiltonian
of Eq.~(33), and where the coefficients are given by
\begin{equation}\label{a} a= \frac{\sum_{j=1}^n\kappa_j/\sqrt{n}}{\sqrt{\sum_{j=1}^n \kappa_j^2}},
\end{equation}
and
\begin{equation}\label{b-term}
b (X_\bot, P_\bot) = \frac{1}{\sqrt{n}} \sum_{i=1}^n  \left(1-
\frac{ \kappa_i \sum_{j=1}^n \kappa_j}{\sum_{j=1}^n \kappa_j^2}
\right)  (x_{\text{at},i}, p_{\text{at}, i}).
\end{equation}
From Eq.~(\ref{decompo}), it follows that the variances of $X$ and
$P$ may be expressed as
\begin{equation}\label{varX}
    \text{Var}(X)=a^2 \text{Var}(X_\text{eff}) +(1-a^2)/2, \end{equation}
and
\begin{equation}\label{varP}
    \text{Var}(P)=a^2 \text{Var}(P_\text{eff}) +(1-a^2)/2,
    \end{equation}
where we have used that $1= a^2 + b^2$ and that the components
$(X_{\bot},P_{\bot})$ are unaffected by measurements, so
$\text{Var}(X_\bot)=\text{Var}(P_\bot) = 1/2$ for all times (if
atomic decay is not taken into account).
\begin{figure}
\begin{center}
\includegraphics[scale=0.45]{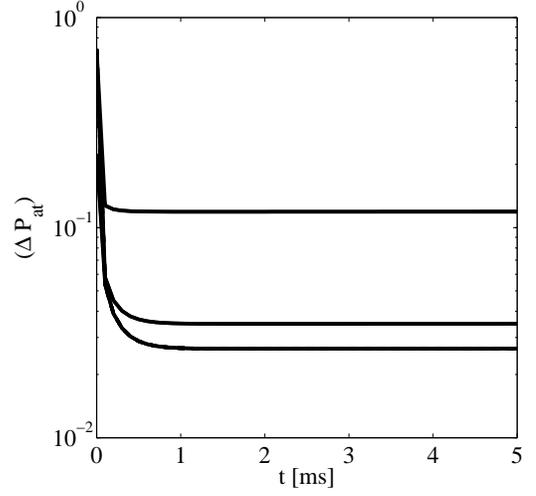}
    \end{center}
    \caption{Uncertainty of the maximally squeezed component of
    the atomic gas ($P_\text{at}$)
    as function of time.
   The higher-lying curves show the uncertainty in the
    symmetric collective parameter (Eq.~(\ref{symm-col-var}))
    for $n=10$
    uniformly distributed values of $\kappa^2$ in
    $[0.9 \kappa_0^2; 1.1 \kappa_0^2]$ (middle)
    and $[0.5 \kappa_0^2; 1.5 \kappa_0^2]$  (upper) .
    The central effective coupling is $\kappa_0^2 =
    1.83 \times 10^{6}$\,s$^{-1}$, and all other parameters are as in Fig.~1.
    The lower curve is the smallest eigenvalue of the covariance
    matrix, which is the same for the two ranges of $\kappa^2$ to
    the precision visible in the figure.
    }
    \label{fig:fig2}
\end{figure}

In Fig.~\ref{fig:fig2}, the lowest curve shows the smallest
eigenvalue of the covariance  matrix as a function of time. The
associated eigenvector represents a combination of the canonical
variables for the different slices, which is maximally squeezed.
For the present values of the noise parameters ($\eta$ and
$\epsilon$), we have an overlap very close to unity between the
eigenvector of this curve and the effective asymmetric collective
variable $P_\text{eff}$ of Eq.~(\ref{x-p-eff}). This means that
this component is indeed the one that is maximally squeezed. The
analytical result for the squeezing of this component is obtained
from (\ref{dpe}) with $\kappa \rightarrow \sqrt{\sum_{j=1}^n
\kappa_j^2}$. For the values for atomic decay and photon
absorption considered in the figure, also the formula (\ref{varP})
reproduces the fully numerical calculations for the symmetric
collective coordinate $P$ of Eq.~(\ref{symm-col-var}).

\subsection{Case (b): optically thick gas}
\label{sec:thick} We now turn to the situation where the sample is
optically thick. The probability $\epsilon$ for absorption of
photons through the gas is then larger than, say, a few percent.
This means that the condition $\epsilon \ll 1$ which was assumed
in the derivation of the effective light-atom coupling of
Eq.~(\ref{hamil3}) is no longer fulfilled. By slicing the gas into
pieces labeled by $i=1,2,\dots,n$, within each of which the
constraint on atomic decay and photon absorption
$\eta_i,\epsilon_i \ll 1$ is fulfilled, we may, however, still
locally for a fixed slice $i$ use the effective Hamiltonian and
address the problem in the gaussian approximation. The vector of
variables describing the system is then of the same form as in
Eq.~(\ref{eq:large-y}), and the Hamiltonian is given by
Eq.~(\ref{Hinhomo}). The considerable absorption of photons from a
beam segment on its way through the atomic gas, means that the
update formula for the covariance matrix needs to be iterated
according to the different local noise and coupling strengths.
Accordingly, as each beam segment passes through the atomic gas
for $i=1$ to $n$, we go through the following update formulae for
the covariance matrix (\ref{gammatrans-withnoise}):
\begin{eqnarray}\label{gammatrans-thick}
\bm{\gamma}_{i} &=& \mathbf{L}_{\tau,i} \mathbf{S}_{\tau,i}
\bm{\gamma}_{i-1} \mathbf{S}^T_{\tau,i} \mathbf{L}_{\tau,i}
\\ \nonumber &+& \frac{\hbar N_{\text{at},i}}{\langle
J_{x,i}(t)\rangle} \mathbf{M}_{\tau,i} + \frac{\hbar
N_{\text{ph},i}}{2 \langle S_{x,i}(t) \rangle} \mathbf{N}_i
\end{eqnarray}
where the transformation matrix $\mathbf{S}_{\tau,i}$ is given by
a matrix with off-diagonal elements $\kappa_{\tau,i}$ at entrances
$((2i-1),(2n+2))$ and $((2n+1),2i))$. For example,
$\mathbf{S}_{\tau,2}$ for the case of only two slices ($n=2$) is
given by
\begin{eqnarray}\label{Sepsi} \mathbf{S}_{\tau,2}=
\left(%
\begin{array}{cccccc}
  1 & 0 & 0 & 0 & 0 & 0 \\
  0 & 1 & 0 & 0 & 0 & 0 \\
  0 & 0 & 1 & 0 & 0 & \kappa_{\tau,2} \\
  0 & 0 & 0 & 1 & 0 &  0 \\
  0 & 0 & 0 & \kappa_{\tau,2} & 1 & 0 \\
  0 & 0 & 0 &  0 &  0 & 1 \\
\end{array}%
\right).
\end{eqnarray}
The constraints on decay  and absorption must be fulfilled
$\eta_{\tau,i}, \epsilon_i \ll 1$ and $\mathbf{L}_{\tau,i}
=\text{diag}(1,\dots,\sqrt{1-\eta_{\tau,i}},
\sqrt{1-\eta_{\tau,i}},1,\dots,\sqrt{1-\epsilon_i},
\sqrt{1-\epsilon_i})$, $\mathbf{M}_{\tau,i}=
\text{diag}(0,\dots,0,\eta_{\tau,i},\eta_{\tau,i},0,\dots,0))$.,
and $\mathbf{N}_i = \text{diag}(0,\dots,0,\epsilon_i,\epsilon_i)$.
The full covariance matrix is updated every time the pulse segment
passes a new slice. When the pulse segment has finally left the
gas, it is being measured, and $\bm{\gamma}_n$ is modified
($\bm{\gamma}_n \rightarrow \bm{\gamma}_n')$ according to
Eqs.~(18) and (19) of Sec.~III, with the $2n \times 2n$ submatrix
$\mathbf{A_\gamma}$ the covariance matrix for the variables
$\bm{y}_1 =
(x_{\text{at},1},p_{\text{at},1},\dots,x_{\text{at},n},p_{\text{at},n})^T$,
$\mathbf{B_\gamma}$ the $2 \times 2$ covariance matrix for
$\bm{y}_2 = (x_{\text{ph}},p_\text{ph})^T$ and $\mathbf{C_\gamma}$
the $2n \times 2$ correlation matrix for $\bm{ y}_1$ and
$\bm{y}_2^T$. When we set $\bm{\gamma}_0(t+\tau) =
\bm{\gamma}_n'(t)$, we use Eq.~(\ref{gammatrans-thick}) with $i=1$
to $n$ to describe the interaction with the next beam segment. In
reality, the light segment corresponding to any practical duration
$\tau$ will be much longer than the entire atomic sample, and the
interaction with one group of atoms has not finished before the
interaction with the subsequent group starts. It is not difficult
to see, however, that if the atomic dynamics is entirely due to
the interaction with the optical field, there is no difference
between the achievements of the real system and those where we
imagine the atomic slices separated by free space separation
distances larger than $c \tau$, described precisely by the above
formulation. In Eq.~(\ref{gammatrans-thick}),

For convenience, we give the time and space (slice) dependence of
the parameters in Eqs.~(\ref{gammatrans-thick}) and (\ref{Sepsi})
explicitly. The change in the classical Stokes vector through the
different slices due to photon absorption is given by
\begin{equation}\label{Sxi}
    \langle S_{x,i} \rangle = \langle S_{x,i=0} \rangle  \exp(-\sum_{i'=1}^i\epsilon_{i'}
    ),
\end{equation}
where the absorption probability in slice $i$ is $\epsilon_i$, and
hence the total photon absorption probability in the gas is
$(1-\exp(-\sum_{i=1}^n \epsilon_i))$. The change in $\langle
J_{x,i} \rangle$ due to atomic decay is given by
\begin{equation}\label{Jxi-t}
    \langle J_{x,i}(t) \rangle = \langle J_{x,i}(0) \rangle  \exp(-\eta_i t).
\end{equation}
The atomic decay rate $\eta_i$ is a decreasing function of the
slice-number since fewer and fewer photons are available to excite
the atoms
\begin{equation}\label{eta-i}
\eta_i = \eta_0\exp ( - \sum_{i'=1}^i\epsilon_{i'}),
\end{equation}
and finally, the light-atom coupling constant $\kappa$ will depend on both time and space
\begin{equation}\label{kappasq-ti}
\kappa^2(t,i) = \kappa_0^2\exp(-\sum_{i'=1}^i \epsilon_{i'})
\exp(- \eta_i t),
 \end{equation} where $\kappa_0^2$ is given as in
Eq.~(\ref{kappa0}) and every slice contains $N_{\text{at},i} =
N_\text{at}$ atoms. From the above relations and the initial
conditions $\langle S_{x,i} \rangle = \hbar N_{\text{ph},i}/2$ and
$\langle J_{x,i} \rangle = \hbar N_{\text{at},i}/2$ it follows
that the pre-factors on the noise terms in
Eq.~(\ref{gammatrans-thick}) are given by
\begin{equation}
\label{prefac-at}
\frac{\hbar N_{\text{at},i}}{\langle J_{x,i}
\rangle} = \frac{2}{e^{-\eta_i t}},
\end{equation}
and
\begin{equation}\label{prefac-ph}
    \frac{\hbar N_{\text{ph},i}}{2 \langle S_{x,i} \rangle } =
    \frac{1}{e^{-\sum_{i'=1}^i \epsilon_{i'} }}.
\end{equation}

\begin{figure} \begin{center} \includegraphics[scale=0.45]{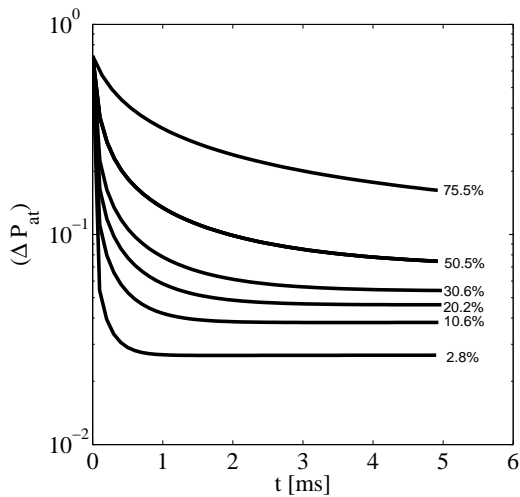}
    \end{center}
    \caption{Uncertainty of the maximally squeezed component of the gas
    ($P_\text{at}$)
    as function of probing time for varying degrees of photon absorption.
    The percentage of photons absorbed is indicated at the solid
    curves. The gas is decomposed in $n$ slices each absorbing
    $\epsilon_i = 0.028$ of the light intensity.
    From the lower to the
    upper curve  the number of such slices attains the values $n=1,4,8,13,25$, and 50.
    Other physical parameters are as specified in Fig. 1.}
    \label{fig:fig3}
\end{figure}
\begin{figure}
\begin{center}
\includegraphics[scale=0.45]{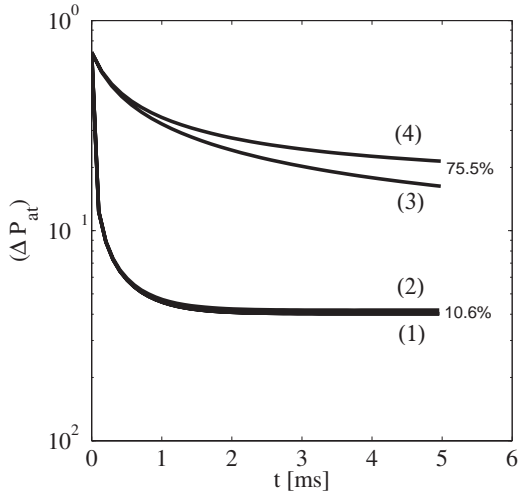}
    \end{center}
    \caption{As Fig.~\ref{fig:fig3} but only for $n=4$ and $n = 50$.
    Curves (1) and (3) represent the maximally squeezed component of the gas. Curves (2) and (4)
    display the uncertainty in the
    inhomogeneous collective variable $P_\text{eff}$ of
    Eq.~(\ref{x-p-eff}) as a function of time. }
    \label{fig:fig4}
\end{figure}

We have modeled the effect of photon-absorption induced
inhomogeneous light-atom coupling using the parameters detailed in
the caption of Fig.~\ref{fig:fig3}. The photon absorption  is
varied by varying the detuning, and the light-atom coupling
strength $\kappa^2$ and the atomic decay probability $\eta$ are
kept constant at the values used in Figs.~\ref{fig:fig1} and
\ref{fig:fig2} by adjusting the photon flux inversely proportional
to changes in the detuning squared. Figure \ref{fig:fig3} shows
the uncertainty of the maximally squeezed component of the sample
as determined by the smallest eigenvalue of the covariance matrix.
We see as expected that the degree of squeezing decreases with
increasing photon absorption probability.

In Fig.~\ref{fig:fig4}, we compare for two representative cases
from Fig.~\ref{fig:fig3}, the uncertainty of the maximally
squeezed component of the gas with the uncertainty of the
collective inhomogeneous variable $P_\text{eff}$  of
Eq.~(\ref{x-p-eff}). The variance of the latter variable can be
calculated straightforwardly from our knowledge of the
time-dependent light-atom coupling constants $\kappa_i$ and the
full covariance matrix: $\text{Var}(P_\text{eff}) =
\left(\sum_{i,j} \kappa_i(t) \kappa_j(t) ( \langle p_i p_j \rangle
- \langle p_i \rangle \langle p_j \rangle ) \right) \big/ \sum_k
\kappa_i(t)^2$. We see that for low and moderate photon
absorption, the result for the effective asymmetric variable of
Eq.~(\ref{x-p-eff}) is close to the fully numerical result. Only
for high photon absorption the effects of noise and differences in
coupling strength lead to a significant deviation from the
numerical result.

\section{Probing the degree of squeezing}
\label{sec:probing}
So far, we have not discussed to which extent
the maximally squeezed component of the atomic sample will be
useful and, e.g., set the limit for the precision obtained in a
measurement of an interesting physical quantity. To investigate
this point, we follow the work in Ref.~\cite{Kuzmich04}, and
consider a situation where (i) the sample is spin squeezed for a
time periode $t_1$ (ii) the spin squeezing is stopped, and the
sample is subject to a spin rotation, and (iii) the system is
probed, and the rotation angle is determined.

\subsection{Noiseless case: Analytical results} \label{sec:prob-an}

We start  by an analysis of the simple case corresponding to a
single atomic sample and a single probe field in the noise-less
limit. From Sec.~\ref{sec:dyn}, we have at time $t_1$
\begin{equation}\label{eq:pat1}
    \text{Var}(p_\text{at}(t_1))  = \frac{1}{2 \kappa^2t_1 +2},
\end{equation}
where we have used that the atoms are initially in a coherent
state with variance $1/2$. Since
$\text{Var}(x_\text{at})\text{Var}(p_\text{at}) = 1/4$ in this
noiseless case, we also have
\begin{equation}\label{eq:xat1}
    \text{Var}(x_\text{at}(t_1))= \kappa^2t_1 /2 + 1/2.
\end{equation}

After the time $t_1$, the light-atom coupling is turned off, and
the system is subject to a rotation around the $y$ axis, described
by the interaction  $H T= - \theta J_y$, where $\theta = \omega T$
is the small angle of rotation resulting from the action of the
constant rotation frequency $\omega$ in time $T$, and where $J_y$
is the $y$ component of the collective spin operator. Making the
translation to the effective dimensionless position operator as in
Eq.~(\ref{eff-at}) leads to the Hamiltonian
\begin{equation}\label{eq:rotation}
    HT = - \hbar \theta \alpha x_\text{at},
\end{equation}
where $\alpha= \sqrt{\langle J_x \rangle /\hbar} =
\sqrt{\frac{N_\text{at}}{2}}$. To obtain an estimate for the
unknown classical variable $\theta$, we follow the ideas
introduced in Ref.~\cite{Moelmer04}, and treat the rotation
variable $\theta$ as a quantum variable within our gaussian
description. The total system is then described by two atomic
variables and one rotation variable
$\bm{y}=(\theta,x_{at},p_{at})^T$. The corresponding
transformation matrix follows from Heisenbergs equations of motion
with the Hamiltonian in Eq.~(\ref{eq:rotation}) and in the basis
($\theta, x_\text{at}, p_\text{at}$) we obtain
\begin{eqnarray}\label{eq:S-rot}
\mathbf{S} =
\left(%
\begin{array}{ccc}
  1 & 0 & 0 \\
  0 & 1 & 0 \\
  \alpha & 0 & 1 \\
\end{array}
\right).
\end{eqnarray}
from which we verify that, e.g, $p_\text{at} \rightarrow
p_\text{at} + \alpha \theta$.  Equation
(\ref{gammatrans-withoutnoise}) now determines the time-evolution of the system,
and we find the following covariance matrix at time $t_2$ after the rotation:
\begin{widetext}
\begin{eqnarray} \label{eq:gammarot-t2} \mathbf{\gamma}(t_2) =
\left(%
\begin{array}{ccc}
  2\text{Var}(\theta_0) & 0 & \alpha 2\text{Var}(\theta_0) \\
  0 & 2 \text{Var}(x_\text{at}(t_1)) & 0 \\
  \alpha 2\text{Var}(\theta_0)& 0 & 2 \text{Var}(p_\text{at}(t_1))+
  \alpha^2 2 \text{Var}(\theta_0) \\
\end{array}%
\right),
\end{eqnarray}
\end{widetext}
where $\text{Var}(p_\text{at}(t_1)) $ and $\text{Var}(
x_\text{at}(t_1))$ are given by Eqs.~(\ref{eq:pat1}) and
(\ref{eq:xat1}), respectively.

Finally, at times $t \ge t_2$, the rotation is turned off, and the
sample is probed by the light beam as in the time interval $[ 0;
t_1]$. The transformation matrix is determined by Heisenberg's
equations of motion for the variables $\bm{y}=(\theta,
x_\text{at}, p_\text{at}, x_\text{ph}, p_\text{ph})^T$ with the
Hamiltonian (\ref{hamil3}) and is given by
\begin{eqnarray} \mathbf{S} =
\left(%
\begin{array}{ccccc}
  1 & 0 & 0 & 0 & 0 \\
  0 & 1 & 0 & 0 & \kappa \\
  0 & 0 & 1 & 0 & 0 \\
  0 & 0 & \kappa & 1 & 0 \\
  0 & 0 & 0 & 0 & 0 \\
\end{array}%
\right).
\end{eqnarray}
The covariance matrix of the system is propagated according to
Eq.~(\ref{gammatrans-withoutnoise}). The measurements on the
photon field are described as in Eqs.~(\ref{gammadecompose}) and
(\ref{Agammaupdate}) (see also Ref.~\cite{Moelmer04}). The
submatrix $\mathbf{A}_\gamma$ is now the $3 \times 3$ matrix
pertaining to the variables ($\theta, x_\text{at},p_\text{at}$),
and $\mathbf{C}_\gamma$ is the  $3 \times 2$ covariance submatrix
describing the coherences and correlations between these three
variables and the photon field. We are interested in the
uncertainty on the value of $\theta$, i.e., the (1,1) entrance in
the covariance matrix. To find this as a function of time, we
follow the procedure in Sec.~\ref{sec:dyn} and calculate the
difference between $\mathbf{A}_\gamma$ after $n$ and $n+1$
iterations, and consider the limit of infinitesimal time steps. In
general, the differential equations obtained in this way are
matrix Ricatti equations and may be solved in standard ways
\cite{stockton}. In the present case, the solution reads for
probing times $t \ge t_2$:
\begin{widetext}
\begin{equation}\label{vartheta}
    \text{Var}(\theta(t))  =
    \text{Var}(\theta_0) -
\frac{\text{Re}(\langle(\theta- \langle \theta
\rangle)(p_{at}-\langle p_{at}\rangle)
    \rangle_{t_2})}
    {\text{Var}(p(t_2))} \left(1 -
    \frac{1}{\left(1 + 2\text{Var}(p(t_2)) \kappa^2
    (t-t_2)\right)}\right),
\end{equation}
\end{widetext}
where the covariances at time $t_2$ are given in
Eq.~(\ref{eq:gammarot-t2}). We see from Eq.~(\ref{vartheta}) that
the variance of the variable $\theta$ does not decrease forever.
In the long time limit, we find
\begin{equation}\label{vartheta-longtime} \text{Var}(\theta (t
\rightarrow \infty) ) = \text{Var}(\theta_0)
\left(\frac{\text{Var}
(p_\text{at}(t_1))}{\text{Var}(p_\text{at}(t_1)) +
\alpha^2\text{Var}(\theta_0)} \right).
\end{equation}
This shows, as expected, that the limiting value only depends on
the squeezing and the rotation until time $t_2$. For large
$\alpha$ parameter (many atoms) and for a sufficiently large
initial variance of $\theta$, the result in
Eq.~(\ref{vartheta-longtime}) reduces to
\begin{equation}
\label{vartheta-simple} \text{Var}(\theta) \simeq \frac{\text{Var}(p_\text{at}(t_1))}{\alpha^2}.
\end{equation}
The ratio of the variances of $\theta$ in a measurement with (S)
and without (NS) spin squeezing is given by
\begin{equation}\label{gain}
\frac{\text{Var}(\theta^{\text{S}})}{\text{Var}
(\theta^\text{NS})} = 2 \text{Var}(p^\text{S}_\text{at}(t_1))
\end{equation}
Since $\text{Var}(p_\text{at}(t_1)) \in ] 0; 1/2]$ this shows that
one may gain a significant factor in precision on the variable
$\theta$ by pre-squeezing the sample.

Finally, we note that the result of Eq.~(\ref{vartheta-longtime})
may be obtained directly by considering the corresponding
classical gaussian probability distribution $P(p_\text{at},\theta)
\propto \exp \left( -p_\text{at}^2/(2\text{Var}(p_\text{at})) -
\theta^2/(2 \text{Var}(\theta)) \right)$. As a consequence of the
rotation, $p_\text{at}$ transforms according to $p_\text{at}
\rightarrow p_\text{at} + \alpha \theta$, and therefore the
probability distribution after rotation  reads
$P(p_\text{at},\theta) \propto \exp \left( -(p_\text{at}- \alpha
\theta)^2/(2\text{Var}(p_\text{at})) - \theta^2/(2
\text{Var}(\theta)) \right)$. A measurement of the variable
$p_\text{at}$ leads to a distribution in $\theta$ only,  from
which the variance of $\theta$ is read off with the result given
in Eq.~(\ref{vartheta-longtime}).

\subsection{Noise included: Numerical results}
\label{sec:prob-num} Whereas in Sec.~\ref{sec:prob-an} it is clear
that it is the collective variable $p_\text{at}$ that is squeezed,
in the case of an atomic ensemble with an inhomogeneous light-atom
coupling we only know from the analysis of Secs.~\ref{sec:thin}
and \ref{sec:thick} that there {\it exists} a component that is
squeezed, and that this component for moderate noise is very
accurately approximated by the asymmetric collective variable
$P_\text{eff}$ of Eq.~(\ref{x-p-eff}). The question we address now
is whether it is the variance  of this component that will show up
in a measurement of a classical parameter, such as the rotation
parameter $\theta$.

The formalism necessary for handling this problem was developed in
Secs.~\ref{sec:dyn} and \ref{sec:thick}. In short, for $n$ slices
of gas each fulfilling $\epsilon_i , \eta_i \ll 1$,
($i=1,\dots,n$) we first propagate and perform measurements on the
system of $2n$ collective atomic position and momentum variables
and 2 collective photon position and momentum  variables. At time
$t_1$, the light field is turned off, and the atomic sample is for
$t \in [ t_1; t_2]$ subject to a rotation around the $y$ axis
described by the effective Hamiltonian
\begin{equation}\label{hrot-slices}
    H\tau =-\hbar \theta \sum_i^n \alpha_i x_{\text{at},i},
\end{equation}
with $\theta = \omega T$ as in Sec.~\ref{sec:prob-an} and with
coupling constants $\alpha_i$ determined by a generalization of
the result in Eq.~(\ref{eq:rotation}) \begin{equation}
\label{alpha-num} \alpha_i= \sqrt{\frac{\langle J_{x,i}
\rangle}{\hbar}} = \sqrt{\frac{N_{\text{at},i}}{2} e^{-\eta_i
t_1}}. \end{equation} Spontaneous emission of photons is neglected
in our approach, so the $\alpha_i$'s are fixed by their values at
the instant of time $t_1$  when the photon field is switched off,
and the possibility for stimulated atomic decay disappears. The
transformation matrix $\mathbf{S}$ corresponding to the
Hamiltonian in Eq.~(\ref{hrot-slices}) is readily found from
Heisenberg's equations of motion for the variables ($\theta,
x_{\text{at},1},p_{\text{at},1},\dots,x_{\text{at},n},p_{\text{at},n}$).
Its diagonal entries are unity, the $i=1,\dots,n$ $(\theta,
p_{\text{at},i})$ entries are assigned the values $\alpha_i$, and
the rest are zero; a natural generalization of
Eq.~(\ref{eq:S-rot}). The propagation in time of the covariance
matrix is then determined by Eq.~(\ref{gammatrans-withoutnoise}).
At time $t_2$ the rotation is stopped, and for times $t > t_2$,
the atom-light Hamiltonian is turned on again. First the initial
covariance matrix for $\theta$, atomic slices and the photon field
($\theta,
x_{\text{at},1},p_{\text{at},1},\dots,x_{\text{at},n},p_{\text{at},n},x_\text{ph},p_\text{ph}$)
is set up. This involves the covariance from the previous part
supplemented by the position and momentum variables of the photon
field. The dynamics of this enlarged covariance matrix is
described by suitable modified versions of Eqs.~(17) and (19) of
Sec.~III.

propagation with this enlarged covariance matrix is performed by
the standard equation of Sec.~\ref{sec:dyn} properly adjusting the
transformation matrix and the matrices associated with noise.
\begin{figure} \begin{center} \includegraphics[scale=0.45]{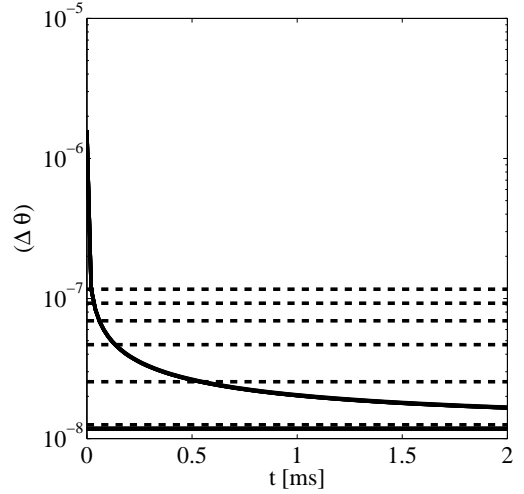}
    \end{center}
    \caption{Uncertainty of the parameter $\theta$ as a function
    of time and for different variances of the coupling strength as specified in the text.
    The gas is sliced in $n=10$
    pieces. The number of atoms and photons are as in the
    preceding figures. The value of $\alpha_i$
    is 0.2236.
    The dashed curves show the limiting uncertainty in the $\theta$
    parameter as estimated from Eq.~(\ref{gen-var-theta}) for the standard collective variable
    $P$ of Eq.~(\ref{symm-col-var}) with the smallest variance in the coupling strength for
    the lowest dashed curve and the highest variance for the upper dashed
    curve.
    The constant horizontal solid line gives the limiting value of
    Eq.~(64), as obtained by the maximally squeezed component $P_\text{eff}$ of
    Eq.~(\ref{x-p-eff}), and it is independent of the variance of the coupling strength.
    The decreasing full curve is a collection of indistinguishable curves
    showing the numerical results for all the different
    variances of the coupling strength (see text).}
    \label{fig:fig5}
\end{figure}

We aim to extract from our numerical study that the variable of
relevance in the probing of the rotation angle is the maximally
squeezed component, i.e., at moderate noise levels it is
essentially the optimally squeezed asymmetric $P_\text{eff}$
variable of Eq.~(\ref{x-p-eff}) and not the symmetric collective
variable $P$ of Eq.~(\ref{symm-col-var}). From Heisenberg's
equation of motion it follows that $P_\text{eff}$ and $P$
transform according to
\begin{equation} \label{theta-transpeff}
P_\text{eff} \rightarrow P_\text{eff} + \left(\frac{\sum_{j=1}^n
\kappa_j \alpha_j}{\sqrt{\sum_{j=1}^n \kappa_j^2}} \right) \theta,
\end{equation}
and
\begin{equation} \label{theta-transp}
P \rightarrow P + \left(\frac{\sum_{j=1}^n \alpha_j}{\sqrt{n}}
\right) \theta.
\end{equation}
A generalization of the result in
Eq.~(\ref{vartheta-simple}) then yields the following expressions
for the variance of $\theta$ in the long time limit
\begin{equation}\label{gen-var-theta-eff}
    \text{Var}(\theta) = \frac{\text{Var}(P_\text{eff})}
    {\left(\frac{\sum_{j=1}^n \kappa_j \alpha_j}
    {\sqrt{\sum_{j=1}^n \kappa_j^2}} \right)^2},
\end{equation}
and
\begin{equation}\label{gen-var-theta}
\text{Var}(\theta') = \frac{\text{Var}(P)}
    {\left(\frac{\sum_{j=1}^n \alpha_j}{\sqrt{n
}} \right)^2},
\end{equation}
where $\text{Var}(P)$ is given by Eq.~(\ref{varP}).

Figure \ref{fig:fig5} shows results for inhomogeneous coupling
modeled by choosing $n=10$ different values of $\kappa^2$
uniformly over the interval $[(1-\delta) \kappa_0^2/n; (1+\delta)
\kappa_0^2/n]$ with $\delta \in \{0,0.02,0.1,0.2,0.3,0.4,0.5\}$.
As in Sec.~\ref{sec:inhom}, the effective coupling strength is
fixed by $\kappa_0$. In the figure, the solid lines are {\it
independent} of fluctuations in the coupling strength. The lowest
solid line shows the asymptotic uncertainty of $\theta$ as
obtained by Eq.~(\ref{gen-var-theta-eff}). The decreasing solid
curve is the numerical result, converging towards this value. It
represents a collection of indistinguishable curves showing the
numerical results for all the different variances of the coupling
strength.  We observe that the decreasing solid curves show a
better estimation of the rotation angle $\theta$ than the
prediction by the symmetric collective variable shown by the
dashed curves in the figure. The fact that the decreasing full
curves converge to the value determined by the maximally squeezed
component signifies that this indeed sets the limit for the
precision of the measurement.

\section{Conclusions}
\label{sec:conclusions}

In this work we have given a comprehensive account of the theory
of probing and measurements in the gaussian state approximation.
We have followed the ideas of Refs.~\cite{Moelmer04,hammerer}, and
we have  provided a  complete analysis of the method and its
strengths by analyzing in detail the problem of spin squeezing.

The gaussian approximation for the collective quantum parameters
including possibly an external classical parameter allows us to
include the measurement process directly, and to obtain analytical
results in the noise-less case and in the limit of low noise. Also
the theory is readily generalized to handle situations which have
resisted a satisfactory treatment with other theoretical methods.
For example, the case of an optically thick gas with corresponding
inhomogeneous light-atom coupling can be treated and even
understood analytically to a large extent.

We have shown that in the present case of squeezing of the spin of
an atomic ensemble by using a continuous wave coherent light beam,
it is indeed the maximally squeezed component of the atomic gas
that determines the precision with which one can estimate the
value of an external perturbation.

At present, we seek to address a series of other problems in
continuous variable quantum physics including generation and
detection of finite band-width squeezed light and estimation of
time-varying external perturbations.

\section*{Acknowledgements}

L.B.M. is supported by the Danish Natural Science Research Council
(Grant No. 21-03-0163).


\begin{thebibliography}{20}
\expandafter\ifx\csname
natexlab\endcsname\relax\def\natexlab#1{#1}\fi
\expandafter\ifx\csname bibnamefont\endcsname\relax
  \def\bibnamefont#1{#1}\fi
\expandafter\ifx\csname bibfnamefont\endcsname\relax
  \def\bibfnamefont#1{#1}\fi
\expandafter\ifx\csname citenamefont\endcsname\relax
  \def\citenamefont#1{#1}\fi
\expandafter\ifx\csname url\endcsname\relax
  \def\url#1{\texttt{#1}}\fi
\expandafter\ifx\csname
urlprefix\endcsname\relax\def\urlprefix{URL }\fi
\providecommand{\bibinfo}[2]{#2}
\providecommand{\eprint}[2][]{\url{#2}}

\bibitem[{\citenamefont{Geremia et~al.}(2003)\citenamefont{Geremia, Stockton,
  Doherty, and Mabuchi}}]{geremia03}
\bibinfo{author}{\bibfnamefont{J.~M.} \bibnamefont{Geremia}},
  \bibinfo{author}{\bibfnamefont{J.~K.} \bibnamefont{Stockton}},
  \bibinfo{author}{\bibfnamefont{A.~C.} \bibnamefont{Doherty}},
  \bibnamefont{and} \bibinfo{author}{\bibfnamefont{H.}~\bibnamefont{Mabuchi}},
  \bibinfo{journal}{Phys. Rev. Lett.} \textbf{\bibinfo{volume}{91}},
  \bibinfo{pages}{250801} (\bibinfo{year}{2003}).

\bibitem[{\citenamefont{M{\o}lmer and Madsen}(2004)}]{Moelmer04}
\bibinfo{author}{\bibfnamefont{K.}~\bibnamefont{M{\o}lmer}} \bibnamefont{and}
  \bibinfo{author}{\bibfnamefont{L.~B.} \bibnamefont{Madsen}},
  \bibinfo{journal}{quant-ph/0402158}  (\bibinfo{year}{2004}).

\bibitem[{\citenamefont{Duan et~al.}(2000)\citenamefont{Duan, Cirac, Zoller,
  and Polzik}}]{duan00}
\bibinfo{author}{\bibfnamefont{L.~M.} \bibnamefont{Duan}},
  \bibinfo{author}{\bibfnamefont{J.~I.} \bibnamefont{Cirac}},
  \bibinfo{author}{\bibfnamefont{P.}~\bibnamefont{Zoller}}, \bibnamefont{and}
  \bibinfo{author}{\bibfnamefont{E.~S.} \bibnamefont{Polzik}},
  \bibinfo{journal}{Phys. Rev. Lett.} \textbf{\bibinfo{volume}{85}},
  \bibinfo{pages}{5643} (\bibinfo{year}{2000}).

\bibitem[{\citenamefont{Julsgaard et~al.}(2001)\citenamefont{Julsgaard,
  Kozhekin, and Polzik}}]{Julsgaard01}
\bibinfo{author}{\bibfnamefont{B.}~\bibnamefont{Julsgaard}},
  \bibinfo{author}{\bibfnamefont{A.}~\bibnamefont{Kozhekin}}, \bibnamefont{and}
  \bibinfo{author}{\bibfnamefont{E.~S.} \bibnamefont{Polzik}},
  \bibinfo{journal}{Nature} \textbf{\bibinfo{volume}{413}},
  \bibinfo{pages}{400} (\bibinfo{year}{2001}).

\bibitem[{\citenamefont{Kuzmich et~al.}(1998)\citenamefont{Kuzmich, Bigelow,
  and Mandel}}]{Kuzmich98}
\bibinfo{author}{\bibfnamefont{A.}~\bibnamefont{Kuzmich}},
  \bibinfo{author}{\bibfnamefont{N.~P.} \bibnamefont{Bigelow}},
  \bibnamefont{and} \bibinfo{author}{\bibfnamefont{L.}~\bibnamefont{Mandel}},
  \bibinfo{journal}{Europhys. Lett.} \textbf{\bibinfo{volume}{42}},
  \bibinfo{pages}{481} (\bibinfo{year}{1998}).

\bibitem[{\citenamefont{Takahasi et~al.}(1999)\citenamefont{Takahasi, Honda,
  Tanaka, Toyoda, Ishikawa, and Yabuzaki}}]{Takahashi99}
\bibinfo{author}{\bibfnamefont{Y.}~\bibnamefont{Takahasi}},
  \bibinfo{author}{\bibfnamefont{K.}~\bibnamefont{Honda}},
  \bibinfo{author}{\bibfnamefont{N.}~\bibnamefont{Tanaka}},
  \bibinfo{author}{\bibfnamefont{K.}~\bibnamefont{Toyoda}},
  \bibinfo{author}{\bibfnamefont{K.}~\bibnamefont{Ishikawa}}, \bibnamefont{and}
  \bibinfo{author}{\bibfnamefont{T.}~\bibnamefont{Yabuzaki}},
  \bibinfo{journal}{Phys. Rev. A} \textbf{\bibinfo{volume}{60}},
  \bibinfo{pages}{4974} (\bibinfo{year}{1999}).

\bibitem[{\citenamefont{Kuzmich et~al.}(1999)\citenamefont{Kuzmich, Mandel,
  Janis, Young, Ejnisman, and Bigelow}}]{Kuzmich99}
\bibinfo{author}{\bibfnamefont{A.}~\bibnamefont{Kuzmich}},
  \bibinfo{author}{\bibfnamefont{L.}~\bibnamefont{Mandel}},
  \bibinfo{author}{\bibfnamefont{J.}~\bibnamefont{Janis}},
  \bibinfo{author}{\bibfnamefont{Y.~E.} \bibnamefont{Young}},
  \bibinfo{author}{\bibfnamefont{R.}~\bibnamefont{Ejnisman}}, \bibnamefont{and}
  \bibinfo{author}{\bibfnamefont{N.~P.} \bibnamefont{Bigelow}},
  \bibinfo{journal}{Phys. Rev. A} \textbf{\bibinfo{volume}{60}},
  \bibinfo{pages}{2346} (\bibinfo{year}{1999}).

\bibitem[{\citenamefont{Bouchoule and M{\o}lmer}(2002)}]{Bouchoule02}
\bibinfo{author}{\bibfnamefont{I.}~\bibnamefont{Bouchoule}} \bibnamefont{and}
  \bibinfo{author}{\bibfnamefont{K.}~\bibnamefont{M{\o}lmer}},
  \bibinfo{journal}{Phys. Rev. A} \textbf{\bibinfo{volume}{66}},
  \bibinfo{pages}{043811} (\bibinfo{year}{2002}).

\bibitem[{\citenamefont{Muller et~al.}(2004)\citenamefont{Muller, Petrov,
  Oblak, Alzar, de~Echaniz, and Polzik}}]{Muller04}
\bibinfo{author}{\bibfnamefont{J.~H.} \bibnamefont{Muller}},
  \bibinfo{author}{\bibfnamefont{P.}~\bibnamefont{Petrov}},
  \bibinfo{author}{\bibfnamefont{D.}~\bibnamefont{Oblak}},
  \bibinfo{author}{\bibfnamefont{C.~L.~G.} \bibnamefont{Alzar}},
  \bibinfo{author}{\bibfnamefont{S.~R.} \bibnamefont{de~Echaniz}},
  \bibnamefont{and} \bibinfo{author}{\bibfnamefont{E.~S.}
  \bibnamefont{Polzik}}, \bibinfo{journal}{quant-ph/0403138}
  (\bibinfo{year}{2004}).

\bibitem[{\citenamefont{Thomsen et~al.}(2002)\citenamefont{Thomsen, Mancini,
  and Wiseman}}]{Thomsen02}
\bibinfo{author}{\bibfnamefont{L.~K.} \bibnamefont{Thomsen}},
  \bibinfo{author}{\bibfnamefont{S.}~\bibnamefont{Mancini}}, \bibnamefont{and}
  \bibinfo{author}{\bibfnamefont{H.~M.} \bibnamefont{Wiseman}},
  \bibinfo{journal}{J. Phys. B: At. Mol. Opt. Phys.}
  \textbf{\bibinfo{volume}{35}}, \bibinfo{pages}{4937} (\bibinfo{year}{2002}).

\bibitem[{\citenamefont{Kuzmich and Kennedy}(2004)}]{Kuzmich04}
\bibinfo{author}{\bibfnamefont{A.}~\bibnamefont{Kuzmich}} \bibnamefont{and}
  \bibinfo{author}{\bibfnamefont{T.~A.~B.} \bibnamefont{Kennedy}},
  \bibinfo{journal}{Phys. Rev. Lett.} \textbf{\bibinfo{volume}{92}},
  \bibinfo{pages}{030407} (\bibinfo{year}{2004}).

\bibitem[{\citenamefont{Geremia et~al.}(2004)\citenamefont{Geremia, Stockton,
  and Mabuchi}}]{Geremia04}
\bibinfo{author}{\bibfnamefont{J.~M.} \bibnamefont{Geremia}},
  \bibinfo{author}{\bibfnamefont{J.~K.} \bibnamefont{Stockton}},
  \bibnamefont{and} \bibinfo{author}{\bibfnamefont{H.}~\bibnamefont{Mabuchi}},
  \bibinfo{journal}{Science} \textbf{\bibinfo{volume}{304}},
  \bibinfo{pages}{270} (\bibinfo{year}{2004}).

\bibitem[{\citenamefont{Kraus et~al.}(2003)\citenamefont{Kraus, Hammerer,
  Giedke, and Cirac}}]{Kraus03}
\bibinfo{author}{\bibfnamefont{B.}~\bibnamefont{Kraus}},
  \bibinfo{author}{\bibfnamefont{K.}~\bibnamefont{Hammerer}},
  \bibinfo{author}{\bibfnamefont{G.}~\bibnamefont{Giedke}}, \bibnamefont{and}
  \bibinfo{author}{\bibfnamefont{J.~I.} \bibnamefont{Cirac}},
  \bibinfo{journal}{Phys. Rev. A} \textbf{\bibinfo{volume}{67}},
  \bibinfo{pages}{042314} (\bibinfo{year}{2003}).

\bibitem[{\citenamefont{Hammerer et~al.}(2003)\citenamefont{Hammerer,
  M{\o}lmer, Polzik, and Cirac}}]{hammerer}
\bibinfo{author}{\bibfnamefont{K.}~\bibnamefont{Hammerer}},
  \bibinfo{author}{\bibfnamefont{K.}~\bibnamefont{M{\o}lmer}},
  \bibinfo{author}{\bibfnamefont{E.~S.} \bibnamefont{Polzik}},
  \bibnamefont{and} \bibinfo{author}{\bibfnamefont{J.~I.} \bibnamefont{Cirac}},
  \bibinfo{journal}{quant-ph/0312156}  (\bibinfo{year}{2003}).

\bibitem[{\citenamefont{Giedke and Cirac}(2002)}]{GiedkeCirac}
\bibinfo{author}{\bibfnamefont{G.}~\bibnamefont{Giedke}} \bibnamefont{and}
  \bibinfo{author}{\bibfnamefont{J.~I.} \bibnamefont{Cirac}},
  \bibinfo{journal}{Phys. Rev. A} \textbf{\bibinfo{volume}{66}},
  \bibinfo{pages}{032316} (\bibinfo{year}{2002}).

\bibitem[{\citenamefont{Fiur\'{a}\v{s}ek}(2002)}]{Fiurasek02}
\bibinfo{author}{\bibfnamefont{J.}~\bibnamefont{Fiur\'{a}\v{s}ek}},
  \bibinfo{journal}{Phys. Rev. Lett.} \textbf{\bibinfo{volume}{89}},
  \bibinfo{pages}{137904} (\bibinfo{year}{2002}).

\bibitem[{\citenamefont{Eisert and Plenio}(2003)}]{EisertPlenio}
\bibinfo{author}{\bibfnamefont{J.}~\bibnamefont{Eisert}} \bibnamefont{and}
  \bibinfo{author}{\bibfnamefont{M.~B.} \bibnamefont{Plenio}},
  \bibinfo{journal}{International Journal of Quantum Information, Vol. 1, No. 4
  (2003) 479,quant-ph/0312071}  (\bibinfo{year}{2003}).

\bibitem[{\citenamefont{Sherson and M{\o}lmer}(2004)}]{Sherson04}
\bibinfo{author}{\bibfnamefont{J.}~\bibnamefont{Sherson}} \bibnamefont{and}
  \bibinfo{author}{\bibfnamefont{K.}~\bibnamefont{M{\o}lmer}},
  \bibinfo{journal}{in preparation}  (\bibinfo{year}{2004}).

\bibitem[{\citenamefont{Maybeck}(1979)}]{Maybeck}
\bibinfo{author}{\bibfnamefont{P.~S.} \bibnamefont{Maybeck}},
  \emph{\bibinfo{title}{Stochastic Models, Estimation and Control. Volume 1}}
  (\bibinfo{publisher}{Academic Press: New York}, \bibinfo{year}{1979}).

\bibitem[{\citenamefont{Stockton et~al.}(2004)\citenamefont{Stockton, Geremia,
  Doherty, and Mabuchi}}]{stockton}
\bibinfo{author}{\bibfnamefont{J.~K.} \bibnamefont{Stockton}},
  \bibinfo{author}{\bibfnamefont{J.~M.} \bibnamefont{Geremia}},
  \bibinfo{author}{\bibfnamefont{A.~C.} \bibnamefont{Doherty}},
  \bibnamefont{and} \bibinfo{author}{\bibfnamefont{H.}~\bibnamefont{Mabuchi}},
  \bibinfo{journal}{Phys. Rev. A} \textbf{\bibinfo{volume}{69}},
  \bibinfo{pages}{032109} (\bibinfo{year}{2004}).

\end{thebibliography}

\end{document}